\documentclass[apjl]{emulateapj}

\slugcomment{version 2009 Jan 25: fm}

\shorttitle{{\it Chandra} Snapshot Survey of 3C Radio Sources with z$<$0.3}
\shortauthors{Massaro et al.}

\newcommand{\xmm}{{\it XMM-Newton}}

\newcommand{\swf}{{\it Swift}}
\newcommand{\sx}{{\it Swift--XRT}}

\newcommand{\asec}{$^{\prime\prime}$}
\usepackage{apjfonts}

\begin{document}

\title{{\it Chandra} Observations of 3C Radio Sources with z$<$0.3:\\ Nuclei, Diffuse Emission, Jets and Hotspots}

\author{F. Massaro\altaffilmark{1}, 
D.~E.~Harris\altaffilmark{1}
G.~R.~Tremblay\altaffilmark{2}
D.~Axon\altaffilmark{2,11},
S.~A.~Baum\altaffilmark{3}, 
A.~Capetti\altaffilmark{4},
M. Chiaberge\altaffilmark{5,7},
R.~Gilli\altaffilmark{6},
G.~Giovannini\altaffilmark{7,10},
P.~Grandi\altaffilmark{8},
F.~D.~Macchetto\altaffilmark{5},
C.~P.~O'Dea\altaffilmark{2},
G.~Risaliti\altaffilmark{9},
W.~Sparks\altaffilmark{5}
}

\altaffiltext{1}{Harvard, Smithsonian Astrophysical Observatory, 60 Garden Street, Cambridge, MA 02138}
\altaffiltext{2}{Dept of Physics, Rochester Institute of Technology, Carlson Center for Imaging Science 76-3144, 84 Lomb Memorial Dr., Rochester, NY 14623}
\altaffiltext{3}{Carlson Center for Imaging Science 76-3144, 84 Lomb Memorial Dr., Rochester, NY 14623}
\altaffiltext{4}{INAF - Osservatorio Astronomico di Torino, Strada Osservatorio 20, I-10025 Pino Torinese, Italy}
\altaffiltext{5}{Space Telescope Science Institute, 3700 San Martine Drive, Baltimore, MD 21218}
\altaffiltext{6}{INAF - Osservatorio Astronomico di Bologna, Via Ranzani 1, 40127, Bologna, Italy}
\altaffiltext{7}{INAF - Istituto di Radioastronomia di Bologna, via Gobetti 101 40129 Bologna, Italy}
\altaffiltext{8}{INAF-IASF - Istituto di Astrofisica Spaziale e fisica cosmica di Bologna, Via P. Gobetti 101, 40129, Bologna, Italy}
\altaffiltext{9}{INAF - Osservatorio Astronomico di Arcetri, Largo E. Fermi 5, 50125, Firenze, Italy}
\altaffiltext{10}{Dipartimento di Astronomia, Universita' di Bologna, via Ranzani 1, 40127 Bologna, Italy}
\altaffiltext{11}{School of Mathematical \& Physical Sciences, University of Sussex, Falmer, Brighton, BN2 9BH, UK}

\begin{abstract} 
We report on our Chandra Cycle 9 program to observe half of the 60
(unobserved by Chandra) 3C radio sources at z$<$0.3 for 8 ksec each.
Here we give the basic data: the X-ray intensity of the nuclei and any
features associated with radio structures such as hot spots and knots
in jets.  We have measured fluxes in soft, medium and hard bands and
are thus able to isolate sources with significant intrinsic column
density.  For the stronger nuclei, we have applied the standard
spectral analysis which provides the best fit values of X-ray spectral
index and column density.  We find evidence for intrinsic absorption
exceeding a column density of 10$^{22}$ cm$^{-2}$ for one third of our
sources.
\end{abstract}


\section{Introduction}
Extended radio galaxies are classified into two main types (Fanaroff and
Riley 1974; Miley 1980; Bridle 1984). The more powerful sources (FRII)
tend to have an edge-brightened radio structure dominated by compact
bright hot spots.  These sources often show either no or one jet which may
be relativistic along its entire length (Laing 1988; Garrington et al.
1988).  The lower luminosity sources (FRI) tend to have edge-darkened
structures which resemble ``plumes'' and usually exhibit two jets. The
jets in FRIs may initially be launched relativistically but seem to
decelerate on subkpc scales likely through interaction with the
environment (e.g., Laing et al. 2008). There are intrinsic differences
in the central AGN of these two types of sources (e.g., Baum, Zirbel,
O'Dea 1995; Evans et al.  2006, Hardcastle et al. 2009a). Most Narrow
Line FRIIs and a few FRIs show evidence for a hidden quasar continuum
source and Broad Line Region (BLR) (e.g., Cohen et al. 1999; Tadhunter
et al. 2007).  The sources with the hidden quasar also produce optical
emission line nebulae with high ionization lines (High Excitation
Galaxies, HEGs) while those without the hidden quasar produce only low
ionization lines (Low Excitation Galaxies, LEGs) (e.g., Hine and
Longair 1979, Laing et al. 1994, Rector and Stocke 2001). There are also
radio sources with properties intermediate between the FRIs and FRIIs,
e.g., the ``Fat Doubles''  (Owen and Laing 1989).

The two main unsolved issues concern the origin of the FR~I~/~FR~II dichotomy 
(how it is related to different acceleration and emission processes),
and the nature of the different emission line regions between LEGs and HEGs
(see Chiaberge et al. 2002 and Hardcastle et al. 2007).

The morphological features of extragalactic radio sources can be described naturally 
with a small number of components: core, jets, hotspots and lobes.
While their radio to optical emission is typically described in terms of synchrotron radiation
by relativistic particles, the origin of X-ray emission in extended structures (jets and hotspots) 
is still unclear, but certainly non-thermal (Harris \& Krawczynski 2002).
The main open question lies in which mechanism, synchrotron or inverse Compton (IC) scattering,
dominates the X-ray emission. The former describes emission from low power jets (Harris \& Krawczynski 2006), while 
the latter provides a good explanation for high power radio galaxy and quasar jets, 
in which the seed photons for the IC scattering could be the Cosmic Microwave Background (CMB) (Tavecchio et al. 2000).

Only by combining X-ray observations with historical and/or simultaneuos data in other wavebands,
is possible to build up the Spectral Energy Distribution (SED) of cores, jets and 
hotspots and compare them with synchrotron or inverse Compton models to investigate the origin of their emission.

During the last few years several snapshot surveys of 3C radio
galaxies have been carried out using the Hubble Space Telescope in
red, blue, ultraviolet and near-IR continuum and optical spectroscopy
which approaches the statistical completeness of the radio catalog:
$\sim 90$\%.  A ground based spectroscopic program for the whole
sample with the Galileo Telescope has been completed (Buttiglione et al 2009).
We also obtained deep ground based IR $K$-band imaging.

Radio images with arcsec resolution are available for most 3C sources
from colleagues, the NRAO VLA Archive Survey (NVAS), and the archives
of the VLA and MERLIN. VLBA data for some 3C objects with $z<0.2$ have
already been obtained (see e.g. Giovannini et al. 2001, Liuzzo et
al. 2009 and references therein).

Chandra is the only X-ray facility that can offer angular
resolution comparable to the optical and radio. Previous
X-ray studies are mostly biased towards observations
of a special group of X-ray bright sources
or objects with well-known interesting features or peculiarities
instead of carefully selected samples, unbiased with respect to orientation and
spectroscopic classification.

The general goals of our program are to discover
new jets and hot spots, determine their emission processes
on a firm statistical basis, study the nuclear emission of
the host galaxy, and derive spectral energy distributions
(SED) for an unbiased sample of objects. The resulting
dataset will be used to test the unification model and
study the nature of nuclear absorption.

The current paper consists of all the basic data for this Chandra sample.  After
a description of the observations and data reduction (\S\ref{sec:obs}), we give
the general and particular results in \S\ref{sec:results}.  \S\ref{sec:summary}
contains a short summary.

For our numerical results, we use cgs units unless stated otherwise
and we assume a flat cosmology with $H_0=72$ km s$^{-1}$ Mpc$^{-1}$,
$\Omega_{M}=0.27$ and $\Omega_{\Lambda}=0.73$ (Spergel et al., 2007).
Spectral indices, $\alpha$, are defined by flux density,
S$_{\nu}\propto\nu^{-\alpha}$.

\section{Observations and Data Reduction}\label{sec:obs}
The sources observed in this project are listed in Table~\ref{tab:gen}
together with their salient parameters.  Each was observed for a
nominal 8 ksec, and the actual livetimes  are given in
Table~\ref{tab:flux}, together with the nuclear fluxes.  We used the
ACIS-S very faint mode with standard readout times (3.2s).  The 15
observations made through 2008 February had 5 chips turned on: I2, I3,
S1, S2, and S3.  When we analyzed the data from 3C436 taken on 2008
Jan 10, we found a background around twice nominal on the 2 back
illuminated chips.  This had the unfortunate consequence of producing
a total count rate essentially equal to the telemetry saturation of
68.8 c/s.  Therefore, to minimize the chances of this occurring again,
all observations made after 2008 Mar 01 had the back illuminated chip
S1 turned off.  Ex post facto, we checked the evt1 countrate for the
14 other observations made with the 5 chips on.  These ranged from 43
to 61 c/s, safely below the telemetry limit.

The data reduction has been performed following the standard
reduction procedure described in the Chandra Interactive Analysis of
Observations (CIAO) threads
\footnote{($http://cxc.harvard.edu/ciao/guides/index.html$)}, 
using CIAO v3.4 and the Chandra Calibration
Database (CALDB) version 3.4.2.  Level 2
event files were generated using the $acis\_process\_events$ task,
after removing the hot pixels with $acis\_run\_hotpix$.  Events were
filtered for grades 0,2,3,4,6 and we removed pixel randomization.

Lightcurves for every dataset were extracted and checked for high
background intervals; none was found except for 3C 436 which was
contaminated by a very high background during the entire exposure so
that time filtering was not an option.

Astrometric registration was achieved by changing the appropiate keywords in the
fits header so as to align the nuclear X-ray position with that of the radio.
We also registered the HST images in the same way.

\subsection{Fluxmaps}
We created 3 different fluxmaps (soft, medium, hard, in the ranges 0.5 -- 1, 1
-- 2, 2 -- 7 keV, respectively) by dividing the data with monochromatic exposure
maps (with nominal energies of soft=0.8keV, medium=1.4keV, and hard=4keV).  The
exposure maps and the flux maps were regridded to a common pixel size which was
usually 1/4 the size of a natvie ACIS pixel (native=0.492$^{\prime\prime}$).
For sources of large angular extent we used 1/2 or no regridding.  To obtain
maps with brightness units of ergs~cm$^{-2}$~s$^{-1}$~pixel$^{-1}$, we
multiplied each event by the nominal energy of its respective band.

To measure observed fluxes for each feature, we construct an appropriate region (usually
circular) and two adjacent background regions of the same size.
The two background regions were chosen so as to avoid contaminating 
X-ray emission (and also radio emission) and permitted us to
sample both sides of jet features, two areas close to hotspots,
and avoid contamination from weak emission surrounding the nuclei of the galaxies.

We then measure the net flux in each region and in each energy band
with funtools\footnote{http://www.cfa.harvard.edu/~john/funtools}.
Measured fluxes were corrected by the ratio of the mean energy within
the photometric aperture to the nominal energy.  This correction
ranged from a few to 15\%.
A one $\sigma$ error is assigned based
on the usual $\sqrt{number-of-counts}$ in the on and background regions.

\subsection{Spectral Analysis of the stronger nuclei}
\subsubsection{Extraction of Point-Source Nuclear Spectra}
We  have performed  a  ``first pass''  nuclear  point source  spectral
analysis for those  sources with greater than 50  counts in a 2\arcsec
aperture  centered   around  the  nucleus  (see  column   3  of  Table
\ref{tab:flux}).  Sources  with fainter nuclei were  excluded from our
analysis  as  sufficient signal  and  resolution  in  energy space  is
required for  robust constraints  on multi-parameter model  fits.  The
unresolved  nuclear  spectra were  extracted  using  the CIAO  routine
\texttt{psextract}, wherein  sources with  greater than 300  counts in
the 2\arcsec aperture  were binned to 30 counts  while fainter sources
were  not binned. Binned  spectra were  fit using  $\chi^2$ statistics,
while unbinned spectra were  fit using Cash statistics (Cash 1979).

\subsubsection{Model Fits}
We have fit a three-component absorbed powerlaw model to each
extracted spectrum using the NASA HEASARC software package
XSPEC (Arnaud 1996).  The multiplicative model used (in XSPEC syntax) is
\texttt{wabs}$\times$\texttt{zwabs}$\times$\texttt{zpowerlw}, and
consists of a galactic photoelectric neutral hydrogen absorption
component (\texttt{wabs}), a redshifted ``intrinsic'' neutral absorption
component (\texttt{zwabs}), and a redshifted powerlaw
(\texttt{zpowerlw}). The powerlaw itself is defined by the target
redshift, the photon index, and a normalization scale factor.  For
each fit the Milky Way absorption (\texttt{wabs}) was fixed (see
column 3 of Table~\ref{tab:spec}), alongside the target redshift (a
parameter in both \texttt{zwabs} and \texttt{zpowerlw}).  The three
remaining ``thawed'' parameters, consisting of the intrinsic
absorption (hereafter N$_H\left(z\right)$), photon index, and
normalization, were allowed to vary in the ``first pass'' fit.

As  stated  previously  Gaussian  ($\chi^2$) statistics  were  used  in
fitting binned  (to 30  counts) spectra with  greater than  300 counts
(i.e.~yielding at least 10 binned channels in unfiltered energy space)
and Cash  statistics were used  for the fainter unbinned  spectra. The
spectra for  3C~105 and 3C~381 were  binned to 15 counts,  as 30 count
bins sacrificed too much resolution  in energy space, resulting in poor
fits.   All the fitting procedures have been performed in the 0.3- 8.0 
keV energy range.

In general, these first-pass fits with three variable parameters were
unsuccessful in constraining both the photon index and the intrinsic
column density to within a reasonable confidence interval.  An
overriding issue is that, for ostensibly ``heavily absorbed sources''
(judging from hard-to-med band flux ratios), one is limited by steep
absorption falloff at the soft end and by declining effective area on
the hard end, effectively limiting the spectral fit to less than one
decade in energy space.  The first-pass XSPEC fits required fitting
the absorption component and photon index using the same region
of the spectrum, such that typically XSPEC would be successful in
fitting one but yield unphysical or unconstrained values for the other
parameter.
All the results of the spectral analysis are summarized in Tab. 4.

\begin{table*}  
\caption{Source List of the Chandra AO9 Snapshot Survey of  3C Radio
  Sources with z$<$0.3}\label{tab:gen}
\begin{flushleft}
\begin{tabular}{lllllrllllll}
\hline
3C     & Class\tablenotemark{a}  & RA (J2000)   & DEC (J2000)  & z\tablenotemark{b}       & D$_L$
          & scale       & N$_H$\tablenotemark{c}    & m$_v$\tablenotemark{d} & S$_{178}$\tablenotemark{e}  & Chandra    & Obs. Date \\
          &                     & hh mm ss      & dd mm ss         &        &  (Mpc)   & (kpc/'')      & (cm$^{-2}$)&     & (Jy)                 & Obs ID & yyyy-mm-dd \\ 
\hline 
\noalign{\smallskip}
17     &  ? -- BLO     & 00 38 20.528 & -02 07 40.49 & 0.2197 & 1078.8 & 3.516       & 2.86e20 & 18.0 & 20.0 & 9292   & 2008-02-02 \\
18     &  FR II -- BLO & 00 40 50.553 & +10 03 26.78 & 0.188  & 905.7  & 3.111       & 5.33e20 & 18.5 & 19.0 & 9293   & 2008-06-01 \\ 
20     &  FR II -- HEG & 00 43 09.177 & +52 03 36.05 & 0.174  & 831.0  & 2.923       & 1.84e21 & 19.0 & 42.9 & 9294   & 2007-12-31 \\
33.1   &  FR II -- BLO & 01 09 44.237 & +73 11 57.10 & 0.181  & 868.2  & 3.018       & 2.00e21 & 19.5 & 13.0 & 9295   & 2008-04-01 \\
52     &  FR II (X)    & 01 48 28.909 & +53 32 28.04 & 0.2854 & 1454.5 & 4.268       & 1.67e21 & 18.5 & 13.5 & 9296   & 2008-03-26 \\
61.1   &  FR II -- HEG & 02 22 35.571 & +86 19 06.38 & 0.1878 & 904.6  & 3.108       & 7.87e20 & 19.0 & 31.2 & 9297   & 2008-12-05 \\
76.1   &  FR I         & 03 03 15.054 & +16 26 18.83 & 0.0324 & 140.3  & 0.638       & 9.49e20 & 14.9 & 12.2 & 9298   & 2007-12-09 \\
105    &  FR II -- HEG & 04 07 16.453 & +03 42 25.80 & 0.089  & 401.7  & 1.642       & 1.15e21 & 18.5 & 17.8 & 9299   & 2007-12-17 \\
132    &  FR II -- LEG & 04 56 42.919 & +22 49 23.20 & 0.214  & 1047.3 & 3.445       & 2.11e21 & 18.5 & 13.7 & 9329   & 2008-03-26 \\
133    &  FR II -- HEG & 05 02 58.472 & +25 16 25.31 & 0.2775 & 1408.2 & 4.183       & 2.54e21 & 20.0 & 22.3 & 9300   & 2008-04-07 \\ 
135    &  FR II -- HEG & 05 14 08.367 & +00 56 32.28 & 0.1273 & 589.9  & 2.250       & 8.73e20 & 17.1 & 17.3 & 9301   & 2008-01-10 \\ 
153    &  FR II -- LEG & 06 09 32.423 & +48 04 14.64 & 0.2769 & 1404.7 & 4.177       & 1.65e21 & 18.5 & 15.3 & 9302   & 2007-12-07 \\ 
165    &  FR II -- LEG & 06 43 07.400 & +23 19 03.00 & 0.2957 & 1481.6 & 4.317       & 1.93e21 & 19.5 & 13.5 & 9303   & 2008-02-02 \\ 
171    &  FR II -- HEG & 06 55 14.722 & +54 08 57.46 & 0.2384 & 1183.5 & 3.741       & 5.65e20 & 18.9 & 19.5 & 9304   & 2007-12-22 \\ 
184.1  &  FR II -- BLO & 07 43 01.394 & +80 26 26.09 & 0.1182 & 544.4  & 2.111       & 3.15e20 & 17.0 & 13.0 & 9305   & 2008-03-27 \\ 
197.1  &  FR II -- BLO & 08 21 33.605 & +47 02 37.40 & 0.1301 & 604.0  & 2.293       & 4.20e20 & 16.5 &  8.1 & 9306   & 2007-12-16 \\
213.1  &  FR II -- LEG (CSS) & 09 01 05.269 & +29 01 46.88 & 0.1937 & 936.4  & 3.186       & 2.45e20 & 19.0 &  6.6 & 9307   & 2008-04-14 \\
223.1  &  FR II -- HEG (X) & 09 41 24.019 & +39 44 41.62 & 0.1075 & 491.5  & 1.943       & 1.31e20 & 16.4 &  6.0 & 9308   & 2008-01-16 \\
287.1  &  FR II -- BLO & 13 32 53.257 & +02 00 45.60 & 0.2156 & 1056.1 & 3.465       & 1.63e20 & 18.3 &  8.2 & 9309   & 2008-03-23 \\
293    &  FR I  -- LEG & 13 52 17.789 & +31 26 46.44 & 0.045  & 196.7  & 0.873       & 1.27e20 & 14.4 & 12.7 & 9310   & 2008-03-19 \\
300    &  FR II -- HEG & 14 22 59.861 & +19 35 36.72 & 0.27   & 1364.5 & 4.101       & 2.49e20 & 18.0 & 17.9 & 9311   & 2008-03-21 \\
303.1  &  FR II -- HEG (CSS)  & 14 43 14.800 & +77 07 29.00 & 0.267  & 1347.1 & 4.068       & 3.16e20 & 19.0 &  8.1 & 9312   & 2008-02-21 \\ 
305    &  FR I  -- HEG (CSS)  & 14 49 21.661 & +63 16 14.12 & 0.0416 & 181.4  & 0.811       & 1.31e20 & 13.7 & 15.7 & 9330   & 2008-04-07 \\ 
315    &  FR I  -- LEG (X)   & 15 13 40.054 & +26 07 30.06 & 0.1083 & 495.4  & 1.955       & 4.27e20 & 16.3 & 17.8 & 9313   & 2007-12-10 \\ 
323.1  &  Quasar -- BLO & 15 47 43.545 & +20 52 16.54 & 0.2643 & 1331.4 & 4.038       & 3.79e20 & 16.7 &  9.7 & 9314   & 2008-06-01 \\
332    &  FR II -- BLO & 16 17 42.540 & +32 22 34.49 & 0.1515 & 713.2  & 2.608       & 1.79e20 & 16.0 &  9.6 & 9315   & 2007-12-10 \\
349    &  FR II -- HEG & 16 59 28.893 & +47 02 55.04 & 0.205  & 997.8  & 3.332       & 1.88e20 & 19.0 & 13.3 & 9316   & 2008-12-28 \\
381    &  FR II -- HEG & 18 33 46.301 & +47 27 02.61 & 0.1605 & 760.0  & 2.736       & 6.15e20 & 17.5 & 16.6 & 9317   & 2008-02-21 \\
436    &  FR II -- HEG & 21 44 11.743 & +28 10 18.91 & 0.2145 & 1050.0 & 3.451       & 6.42e20 & 18.2 & 17.8 & 9318   & 2008-01-08 \\ 
460    &  FR II -- LEG & 23 21 28.510 & +23 46 48.45 & 0.268  & 1352.9 & 4.079       & 4.72e20 & 18.8 &  8.2 & 9319   & 2008-06-04 \\ 
\noalign{\smallskip}
\hline
\end{tabular}\\
\end{flushleft}
(a) The 'class' column contains both a radio descriptor (Fanaroff-Riley class I or II) and a spectroscopic designation, LEG,
``Low Excitation Galaxy'', HEG, ``High Excitation Galaxy'', and BLO, ``Broad Line Object''; Buttiglione et al. 2009). ``CSS'' designates ``Compact Steep Spectrum'' (a radio term) while ``X'' is used for ``X-shaped'' radio morphology.\\
(b) Redshfit estimates are taken form Chiaberge et al. 2002, Floyd et al. 2006, Buttiglione et al. 2009.\\
(c) Neutral hydrogen column densities are taken form Kalberla et al. 2005.\\
(d) $m_v$ is the visual magnitude (Spinrad et al. 1985).\\
(e) S$_{178}$ is the flux density at 178 MHz, taken from Spinrad et al. 1985.\\
\end{table*}

\section{Results}\label{sec:results}

\subsection{General}
X-ray emission was detected for all the nuclei in the sample except
for 3C 153, a small size FR II radio galaxy with a radio quiet (< 0.5
mJy/beam) core (see Laing et al. 1981 and Fig.~\ref{fig:app153}).  The
observed nuclear fluxes are presented in Table~\ref{tab:flux} in the
soft, medium, and hard bands together with the X-ray luminosity.  The
number of counts and the ratio of net counts (r=2$^{\prime\prime}$
divided by r=10$^{\prime\prime}$) are also given.  This ratio should
be close to unity for an unresolved source: the on-axis encircled
energy for r=2$^{\prime\prime}$ is $\approx$0.97 so we expect only a
small increase between r=2$^{\prime\prime}$ and r=10$^{\prime\prime}$
for an unresolved source.

\begin{figure}
\includegraphics[scale=0.30,origin=c,angle=0]{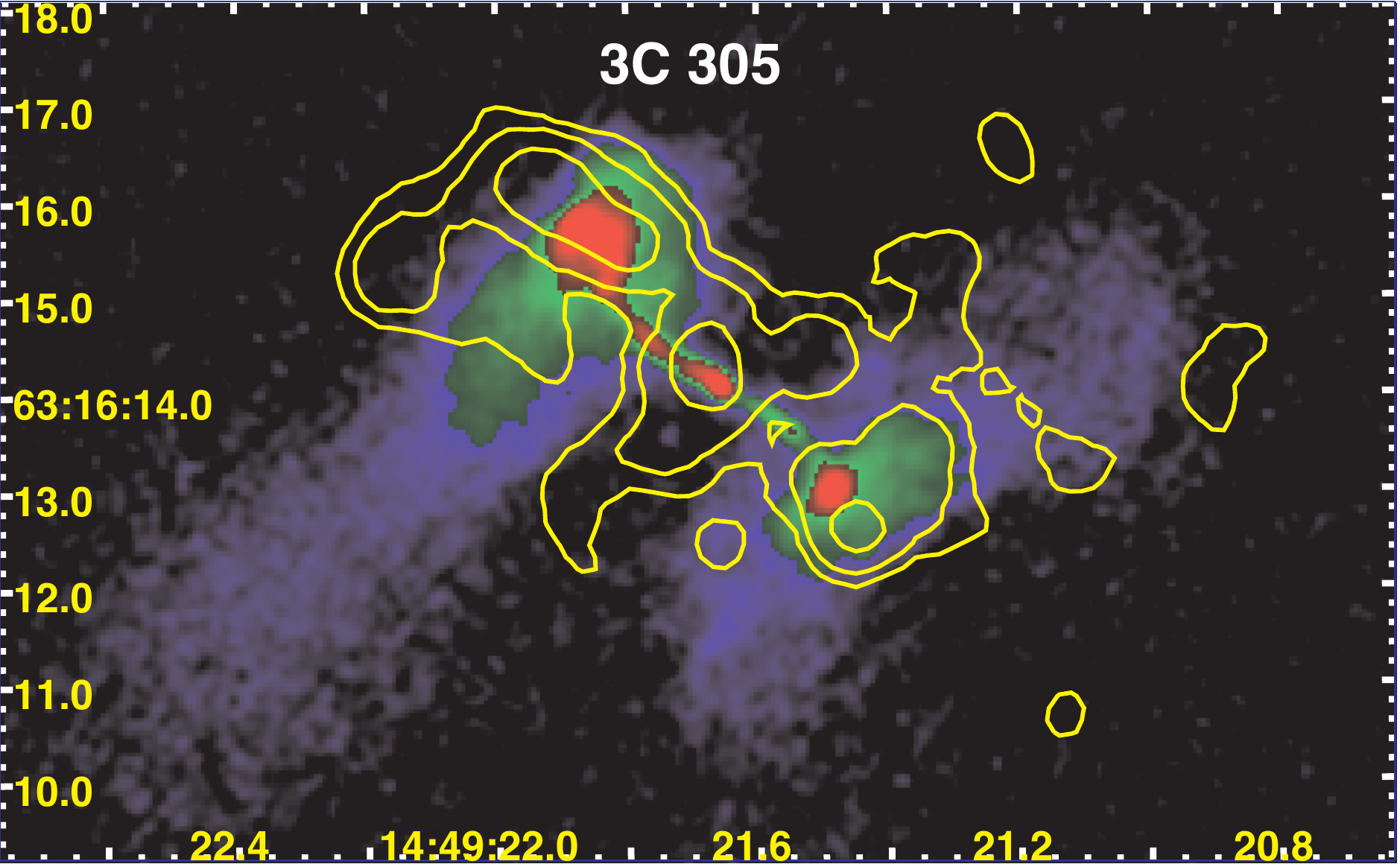}
\caption{3C~305, an example of a CSS source.  The radio image is from
a 1.5 GHz observation (see Massaro et al. 2009a and reference therein).
The X-ray contours come from our 0.5 to 7 keV
image, rebinned to 0.123$^{\prime\prime}$ pixels and smoothed with a
Gaussian of FWHM=0.72$^{\prime\prime}$.  The lowest contour is 0.04
counts per pixel and successive contours increase by factors of two.}
\label{fig:305}
\end{figure}

\begin{figure}
\includegraphics[scale=0.30,origin=c,angle=0]{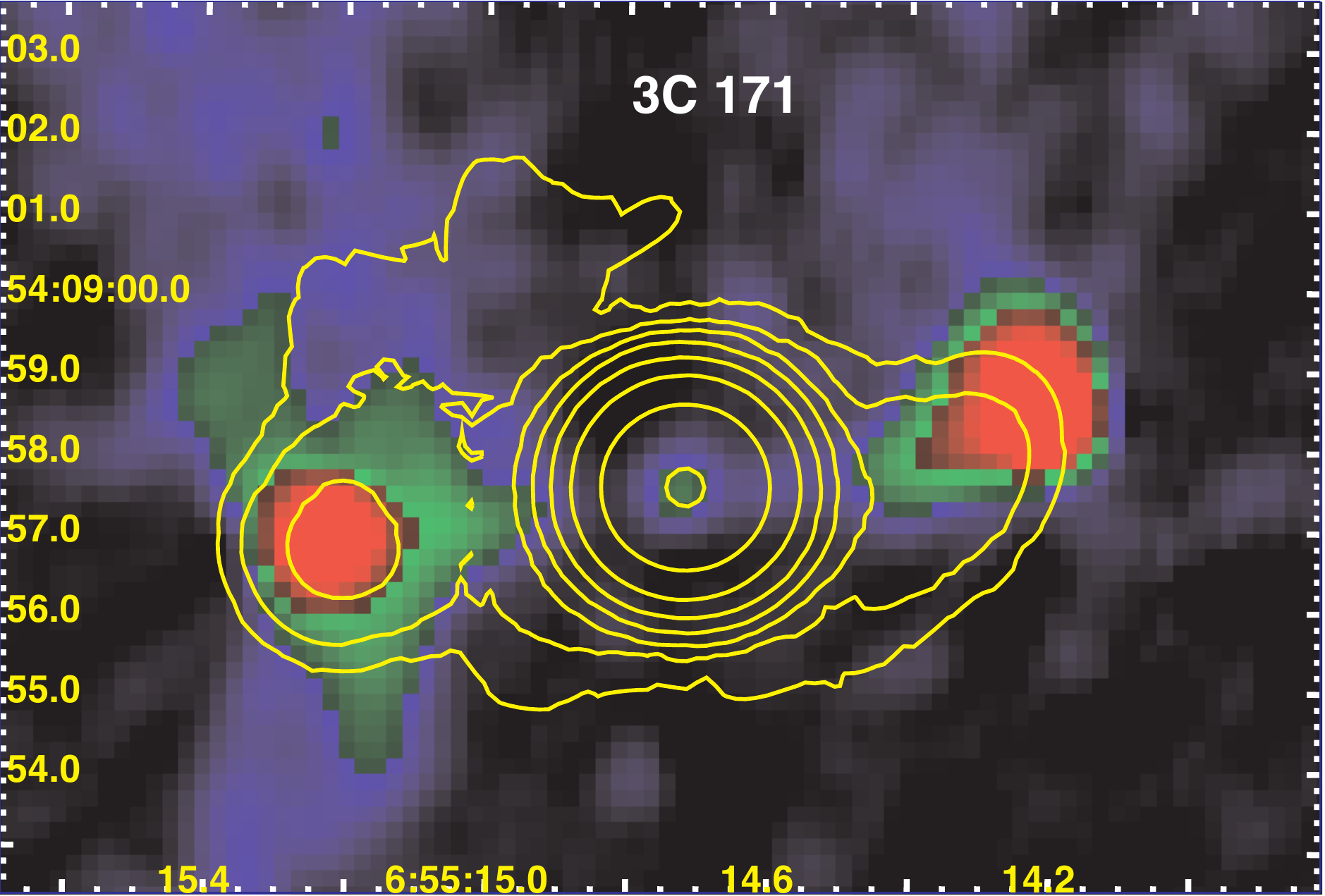}
\caption{3C~171, a small FRII radio galaxy with extended emission
surrounding the nucleus.  The radio image comes from an 8 GHz VLA
observation (Hardcastle 2003).  
The X-ray contours come from an image rebinned as in
Fig.~\ref{fig:305}, smoothed with a Gaussian of FWHM=1.9\asec.  The
lowest contour is at 0.005 counts per pixel and contours increase by
factors of two. (Radio map kindly supplied by M. Hardcastle).}
\label{fig:171}
\end{figure}

\begin{figure}
\includegraphics[keepaspectratio=true,scale=0.40]{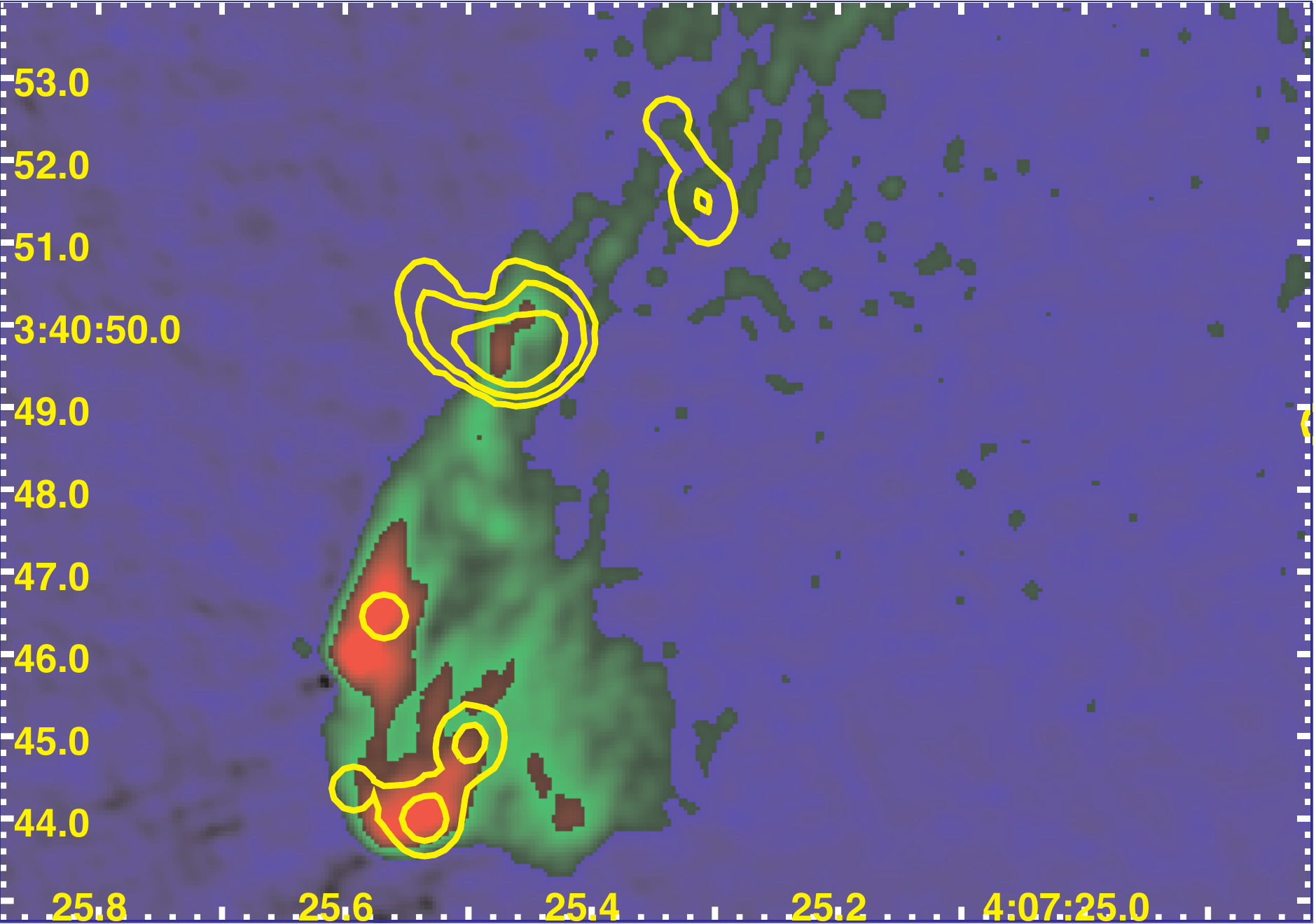}
\caption{\footnotesize The S hotspot of 3C~105.  The image is from the
VLA; an 8 GHz map with a beam of FWHM=0.25$^{\prime\prime}$.  The
contours come from a Chandra fluxmap (0.5-7keV) smoothed with a
Gaussian of FWHM=0.7$^{\prime\prime}$. The lowest contour is 0.02
counts per pixel and the pixel size is 0.123$^{\prime\prime}$.
Subsequent contours increase by factors of two.  Note that the
brighter X-ray emission coincides with a radio feature which could be
described as the jet entering the hotspot region.  X-ray emission is
also detected from the (radio) brighter terminal hotspot.
(Radio map kindly supplied by M. Hardcastle).}
\label{fig:105}
\end{figure}

Amongst the detected sources are 3 compact steep spectrum (CSS) radio
sources: 3C 213.1, 3C 303.1 and 3C 305 (see Fig.~\ref{fig:305}).  A
detailed analysis of 3C 305 has already been published
(Massaro et al. 2009a).  We also found a radio source with soft X-ray
emission presumably associated with thermal gas of the host galaxy, 3C
171 (see Fig.~\ref{fig:171}).  For 6 of our sources we have detections
of a hotspot with confidence levels between 1 and 14 sigma.  In 3C~105
(see Fig.~\ref{fig:105}), we find two emission regions, one coincident
with the radio hotspot itself, and another which seems to be the jet
entering into the hotspot region. Fluxes for these are reported in
Table~\ref{tab:hotspot}.  Finally we detected two knots in the curved
jet of 3C 17 (see Fig.~\ref{fig:17}); details have been published
(Massaro et al. 2009b)

\begin{figure}
\includegraphics[scale=0.30,origin=c,angle=0]{./17body_cmyk.pdf}
\caption{3C~17 with a one sided, curved jet.  The radio image is that
of Morganti et al. (1999) at 5 GHz.  The X-ray contours come from our
image with a smoothing function of 1\asec.  The lowest contour is 0.02
counts per pixel (with pixel=0.123\asec).
(Radio map kindly supplied by R. Morganti).}
\label{fig:17}
\end{figure}

\subsection{Incidence of Intrinsic Absorption}
According to the unified scheme, since a large fraction of our current
sample are FRII radio galaxies, we may expect some sources to
show a significant amount of absorbing material along the line of
sight to the nuclei of these galaxies.  Indeed, we see evidence of
high absorption both from the fluxmap data and the spectral analyses.

Most nuclei of AGN have X-ray spectra which are well described by
power laws with $\alpha_x$ values ranging from 0.5 to 1.5 or
occasionally larger.  For those sources with little intrinsic
absorbing material (i.e. the galactic N$_H$ values are the major
contributors to the total absorption), the expected ratio of hard to
medium flux can be easily calculated based on the chosen energies
defining the relevant bands.  For our standard bands, these ratios run
from 1 to 3, depending of course on the actual value of $\alpha_x$.
Even for extreme values of $\alpha_x$ such as 0 or 2.5, the ratios are
5 and 0.5, respectively.  In Table~\ref{tab:flux} we give the observed
values of this ratio, together with the uncertainties derived from the
errors on the flux values.  About half of our sample have ratios that
are larger than 6, even after allowing them to take on their minimum value
(i.e. value minus error).  We suspect that all of these have
significant intrinsic absorption since, in the absence of such
absorption, $\alpha_x$ would have to have a value less than zero.

A similar method to derive evidences of intrinsic absoprtion
from the flux ratios in different energy ranges
have been used by Alexander et al. (2001, 2002) for large sample of
AGNs.
However, if the sources are Compton thick (i.e. N$_H$ > 10$^{24}$ cm$^{-2}$)
or if the spectrum is inverted (i.e. $\alpha_x$ < 0)
the ratios will not provide a good estimate of absorption
In particular, for Compton thick sources, even if they are very rare
among radio loud AGNs, the spectrum could be dominated by a reflection component
providing a ratio of $\sim$ 5-10 indicating a low intrinsic
absorption even if the source is heavly absorbed.

We note that the values of the neutral absorption estimated using the photometric method and
evaluated from the spectral analysis could be different. 
This happen because the photometric method is based on the assumption
that the intrinsic spectrum is a simple power-law. 
If the spectrum has some features, as for example emission lines,
the estimates of intrinsic N$_H$ could be different, as in the case of 3C 105,
and only the detailed spectral fitting procedure will provide the correct information.

Much the same conclusion can be reached by the difficulty of fitting 
a power law plus absorption to the brighter sources.  The fact that
XSPEC returns small values of  $\alpha_x$ we ascribe to
the situation of a heavily absorbed spectral distribution for which
it is difficult to define the actual power law above 2 keV where the
effective area of the instrument is falling.  What we are dealing with
is an inadequate segment of the spectrum which is essentially flat
or inverted, around the peak in the spectral energy distribution.

\begin{figure}
\includegraphics[scale=0.30,origin=c,angle=0]{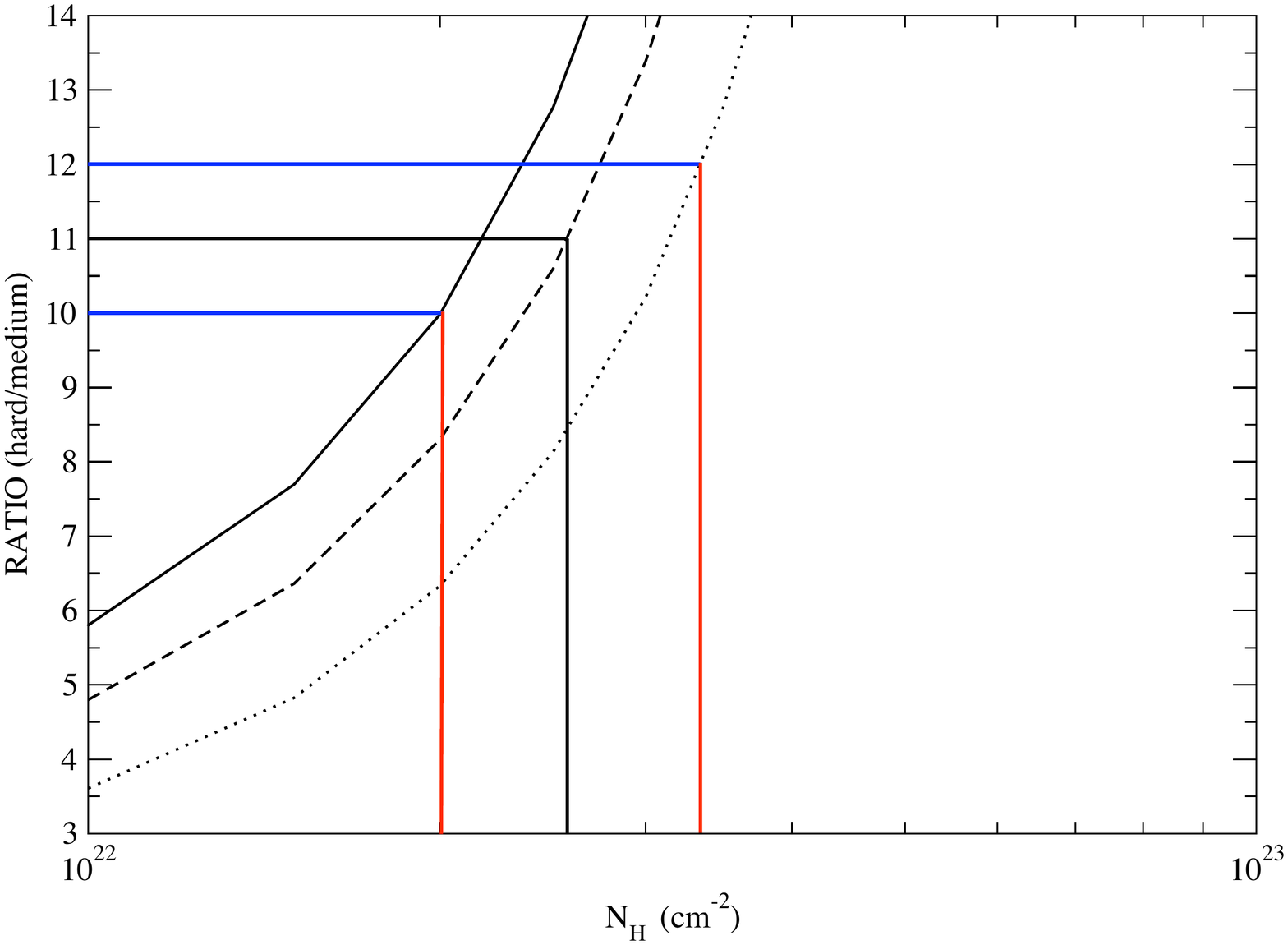}
\caption{Results of simulating spectra for a range of intrinsic
  absorption and a range of $\alpha_x$.  The ratio of hard/medium flux
is plotted on the vertical axis and the column densities are plotted
along the x axis.  The 3 curves (left to right) correspond to
$\alpha_x$ = 0.5, 0.7, and 1.  The example shown is for 3C332 which
has a ratio of 11$\pm$1.}
\label{fig:nh}
\end{figure}

Even though we may not be able to recover the parameters of interest
($\alpha_x$ and N$_H$) from the spectral fits, it is possible to
demonstrate a range of intrinsic N$_H$ column densities corresponding
to some chosen range in $\alpha_x$ by using 'fake' spectra in XSPEC.
An example for 3C 332 is shown in Fig.~\ref{fig:nh}.  Corresponding to
the ratio of hard to medium flux of 11$\pm$1, we find a range of 2 to
3.2$\times10^{22}cm^{-2}$ for $\alpha_x$ between 0.5 and 1.  We have
run these calculations (the ``photometric method'') for each 
source with large hardness ratio and thus suspected to have
large intrinsic absorption. These results are given in
Table~\ref{tab:flux}.  Comparing N$_H$ values from XSPEC to those
calculated from the ratio of hard/medium, we find reasonable agreement
for 5 sources (3C 20, 33.1, 61.1, 171, \& 184.1).  For 3C~133, the
photometric method gives a value 5 times larger than the spectral
analysis, and for 3 sources (3C~105, 223.1, \& 381) we find spectral
values 3 to 5 times larger than their respective photometric values.
Given the difficulties in spectral fitting (\S2,2) together with the
likely wider range in $\alpha_x$ than used in our calculations, we do
not think it necessary to pursue these disparities further.

\begin{table*} 
\caption{Nuclear X-ray Fluxes in units of 10$^{-15}$erg~cm$^{-2}$s$^{-1}$}\label{tab:flux}
\begin{flushleft}
\begin{tabular}{llrlrrrrrrr}
\hline
3C     & LivTim\tablenotemark{a}     &  Net\tablenotemark{b}   &
Ext. Ratio\tablenotemark{c}  & f(soft)  & f(medium) & f(hard) &  f(total)
& $\frac{HARD}{MEDIUM}$\tablenotemark{d} & N$_H^{(e)}$ & L$_X$ \\ 
          & (ksec) & (cnts) &    & 0.5-1~keV & 1-2~keV   & 2-7~keV &
          0.5-7~keV  &  & (10$^{22}$cm$^{-2}$)  & (10$^{42}$erg~s$^{-1}$) \\
\hline 
\noalign{\smallskip}
17     & 7.93 & 1151 (34) & 0.98 (0.04) & 139 (9)     & 193 (9)     & 
695 (34)    &  1030  (37)   & 3.6 (0.2) & ---  & 143  (5)      \\     
18     & 7.94 & 1045 (32) & 0.88 (0.04) & 86.9 (6.9)  & 149 (8)     & 
909 (40)    &  1140  (41)   & 6.1 (0.4) & 1.4 (0.6)  & 112  (4)      \\
20     & 7.93 &  147 (12) & 0.93 (0.12) & ----------- & 1.36 (0.78) & 
286 (24)    &  287   (24)   & 210 (121) & 10.2 (0.7)  & 23.7  (2.0)   \\
33.1   & 8.07 &  692 (26) & 0.95 (0.05) & 3.5 (1.4)   & 40.0 (4.3)  & 
1080 (44)   &  1120  (45)   & 27  (3)    &4.7 (0.7)  & 101  (4)     \\
52     & 8.02 &   11 (3)  & 0.85 (0.60) & ----------- & 1.13 (0.65) & 
14.5 (5.5)  &  15.6  (5.6)  & 12.8  (8.8) & 3.0 (2.0)  & 4.0  (1.4) \\ 
61.1   & 8.05 &   51 (7)  & 0.99 (0.26) & 1.61 (0.93) & 5.9 (1.6) & 
66 (12) &  74  (11) & 11  (3)    & 2.6 (0.7)  & 7.2  (1.1) \\ 
76.1   & 8.06 &   27 (5)  & 0.64 (0.20) & ----------- & 3.1 (1.3) & 
26.2 (6.6)  & 29.3 (6.7)    & 8   (4)    & 1.9 (1.3) & 0.069 (0.016)\\
105    & 8.07 &  308 (18) & 0.93 (0.08) & 3.0 (2.2)   & 6.3 (1.6) & 
699 (41)    &  708   (42)   & 111 (28)   & 8.5 (0.7)  & 13.7 (0.8)   \\ 
132    & 7.69 &   42 (6)  & 1.00 (0.28) & ----------- & 2.3 (1.2) & 
69 (11) &  69  (11) & 30  (16)   & 5.0 (0.6)  & 9.1  (1.5)   \\
133    & 8.03 &  623 (25) & 0.99 (0.06) & 22.7 (3.5)  & 138 (7.7)  & 412 
(26)     &  573   (27)   & 3.0  (1.9)   & ---  & 136  (6)     \\
135    & 7.93 &   40 (6)  & 0.82 (0.21) & 3.7 (1.5)   & 5.0 (1.4) & 
45 (10) &  54  (10) & 9   (3)    & 2.16 (1.19)  & 2.26  (0.43) \\
153    & 8.06 & ---       & ---         & ---         & ---         & 
---         & ---           & ---        & ---  & ---          \\
165    & 7.67 &  35 (6)   & 0.41 (0.10) & ---         & 5.2 (1.5)   & 
43.6 (9.1)  &  48.8 (9.2)   & 8   (3)    & 1.9 (1.3)  & 12.8 (2.4)   \\
171    & 7.93 &  184 (14) & 0.69 (0.07) & 1.7 (1.0)   & 5.6 (1.5) & 
312 (24)    &  319  (24)    & 55  (16)   & 6.5 (0.7)  & 53.5  (4.1)  \\ 
184.1  & 8.02 &  535 (23) & 0.94 (0.06) & 6.6 (1.8)   & 25.6 (3.4)  & 
864 (40)    &  896  (41)    & 34  (4)    & 5.2 (0.7)  & 31.8  (1.4)  \\ 
197.1  & 7.96 &  860 (29) & 0.95 (0.05) & 103 (8)     & 157 (8)     & 
541 (31)    &  801  (33)    & 3.4 (0.3)  & ---  & 35.0  (1.4)  \\ 
213.1  & 8.07 & 65 (8)    & 0.56 (0.10) & 9.5 (2.2)   & 13.6 (2.3) & 
21.2 (6.1) & 44.3 (6.9)    & 1.6 (0.5)  & ---  & 4.65 (0.72)  \\ 
223.1  & 7.93 &  178 (13) & 0.90 (0.10) & 8.6 (2.1)   & 4.1 (1.3) & 
313 (32)    &  326  (32)    & 76  (25)   & 7.4 (0.7)  & 9.42  (0.92) \\
287.1  & 8.02 & 1095 (33) & 0.90 (0.04) & 117 (8)     & 169 (8)     & 
763 (39)    &  1050  (40)   & 4.5 (0.3)  & ---  & 140  (5)     \\ 
293    & 7.81 & 215 (15)  & 0.69 (0.07) & 4.0 (1.5)   & 6.6 (1.7)   & 
381 (28)    & 392 (29)      & 58  (15)   & 6.7 (0.7)  & 1.81 (0.13)  \\ 
300    & 7.94 &  162 (13) & 0.92 (0.11) & 23.6 (3.5)  & 30.7 (3.6)  & 
67 (10) &  122  (11)    & 2.2 (0.4)  & ---  & 27.2  (2.5)  \\
303.1  & 7.67 & 19 (4)    & 0.41 (0.13) & 3.7 (1.4) & 1.87 (0.93) & 
9.9 (4.4) &  15.8 (4.7)   & 5   (4)    & ---  & 3.4 (1.0)  \\ 
305    & 8.22 & 63 (8)    & 0.31 (0.04) & 12.0 (2.5)  & 10.4 (2.0)  & 
13.8 (4.9)  &  36.2 (5.9)   & 1.3 (0.5)  & ---  & 0.143 (0.023)\\ 
315    & 7.67 &    6 (2)  & 0.50 (0.36) & ---         & ---         & 
9.6 (4.8) &  9.6  (4.8) & ---   & ---  & 0.28 (0.14)\\
323.1  & 7.93 &  512 (23) & 0.70 (0.04) & 31.6 (3.9)  & 62.7 (6.3)  & 
532 (35)    &  626  (36)    & 8   (1)    & 1.9 (0.8)  & 133  (8)     \\
332    & 7.93 &  736 (27) & 0.80 (0.04) & 29.3 (3.9)  & 82.7 (6.1)  & 
931 (43)    &  1040  (43) & 11  (1)    & 2.6 (0.7)  & 63.3  (2.6)  \\
349    & 8.02 &  230 (15) & 0.97 (0.10) & 5.4 (1.7) & 40.0 (4.2)  & 
212 (19)    &  257  (19)    & 5.3 (0.7)  & ---  & 30.6  (2.3)  \\ 
381    & 8.06 &  246 (16) & 0.96 (0.10) & 10.3 (2.2)  & 9.4 (2.0) & 
415 (29)    &  435  (30)    & 44  (10)   & 5.9 (0.5)  & 30.1  (2.0)  \\
436    & 8.04 & 43 (7)    & 0.27 (0.06) & 2.4 (1.2) & 2.25 (0.92) & 68 
(12)     &  72 (12)      & 30  (13)   & 5.0 (0.6)  & 9.5 (1.6)    \\ 
460    & 8.05 &   44 (7)  & 0.88 (0.20) & 4.9 (1.6) & 3.2 (1.2) & 
50.4 (9.7)  &  58.5  (9.9) & 16  (7)    & 3.4 (0.5)  & 12.8  (2.2)  \\
\noalign{\smallskip}
\hline
\end{tabular}\\
\end{flushleft}
\tablecomments{Values in parentheses are 1$\sigma$ uncertainties.}

(a) LivTim is the live time\\
(b) Net is the net counts within a circle of radius=
2$^{\prime\prime}$.\\
(c) Ext. Ratio (``Extent Ratio'') is the ratio of the
net counts in the r=2$^{\prime\prime}$ circle to the net counts in the
r=10$^{\prime\prime}$ circle.  Values significantly less than 0.9
indicate the presence of extended emission around the nuclear
component.\\
(d) In the absence of absorption, the ratio of hard flux
to medium flux should lie in the range 0.5 to 5 for 'normal' power
laws (see text).  Thus values significantly larger than 6 indicate
substantial intrinsic absorption.\\
(e) As per the discussion in the text, we calculate the
value of N$_H$ required to produce the observed ratio of hard/medium
flux.  The uncertainty given here is indicative only: it is the range
of N$_H$ covered by the uncertainty in the ratio and allowing
$\alpha_x$ to range from 0.5 to 1.0.  Obviously there may be some
sources with intrinsic spectral indices outside of this range.\\
\end{table*}

\begin{table*}  
\caption{Radio Hotspots with X-ray Detections}\label{tab:hotspot}
\begin{flushleft}
\begin{tabular}{lllllllll}
\hline
3C      & Hotspot\tablenotemark{a}  & Radius\tablenotemark{b} & counts (bkg)\tablenotemark{c}  & f$_{0.5-1~keV}$  & f$_{1-2~keV}$  & f$_{2-7~keV}$   & f$_{0.5-7~keV}$   & L$_X$    \\ 
           &                                  & (arcsec)
           &
           &  (cgs)  & (cgs) &
           (cgs) &
   (cgs)        &10$^{41}$erg~s$^{-1}$ \\
\hline 
\noalign{\smallskip}
52      & N~34      & 1.2  &   5 (0.45) & 1.21 (0.85) & 1.27 (0.75) & -------     & 2.5 (1.1)   & 6.3   (2.9)  \\        
61.1    & S~102    & 4.0  &   9 (5.2)  & 1.56 (0.90) & 1.08 (0.78) & 4.8 (2.8) & 7.5 (4.5)   & 7.3   (4.3)  \\
105     & S~166    & 1.5  &  14 (1.1)  & 2.3 (1.1)   & 1.74 (0.87) & 9.7 (4.0)   & 13.7 (4.2)  & 2.65  (0.82) \\
        & S~169    & 1.5  &   5 (1.1)  & 0.46 (0.46) & 0.52 (0.52) & 4.8 (2.8)   & 5.8 (2.9)   & 1.11  (0.55) \\
213.1 & N~4            & 1.0  &   3 (0.40) &  --------   & 0.70 (0.50) & 0.63 (0.17) & 1.3 (0.5) & 1.40  (0.54) \\
287.1 & W~65          & 5.0  &  10 (7.7)  & 2.3 (1.2)   & 1.30 (0.90) & 4.5 (3.2)   & 8.1 (3.5)   & 10.8  (4.6)  \\
349     & S~38      & 2.0  &   4 (0.70) & 0.51 (0.51) & 0.67 (0.47) & 1.5 (1.5)   & 2.7 (1.6)   & 3.17  (2.0)  \\
        & S~43      & 2.0  &   6 (0.70) & 0.40 (0.40) & 1.55 (0.77) & 0.98 (0.98) & 2.9 (1.3)   & 3.48  (1.56) \\
\noalign{\smallskip}
\hline
\end{tabular}\\
\end{flushleft}
Fluxes are given in units of 10$^{-15}$erg~cm$^{-2}$s$^{-1}$.\\
(a) The hotspot designation is comprised of a cardinal direction plus the distance from the nucleus is arcsec.\\
(b) The radius column gives the size of the aperture used for photometry.\\
(c) The counts column gives the total counts in the photometric circle together with the expected background counts in parantheses; bothfor the 0.5 to 7 keV band.\\
\end{table*}

\begin{table*}        
\caption{Best-fit    Parameters     from    Spectral    Analysis}
\label{tab:spec}
\begin{flushleft}
\begin{tabular}{lrccccc}
\hline
  &  
Bin   Size  &
Galactic N$_H$  & 
Intrinsic N$_H\left(z\right)$  & 
 & 
 &
 \\   
Source  &   
(counts)   &  
($\times 10^{22}$  cm$^{-2}$) &  
($\times 10^{22}$  cm$^{-2}$) &
$\alpha_x$ & 
Red. $\chi^{2}$ &
Cstat/dof(dof)    \\   
(1)   &    (2)   & (3)&   (4)  &  (5)  &   (6)  & (7) \\
\hline 
\noalign{\smallskip}
3C~17  & 30 & 0.0286 & $<0.05$  & 0.3 (0.1) & 0.81       & 0.664 (333) \\
3C~18 & 30 & 0.0533 & $<0.09$ & -0.16 (0.12) &  1.50  & 0.853 (342) \\
3C~20 & 1 & 0.1840 & 13 (6)  & 0.5 (1.1)  & \nodata           & 0.802 (113) \\
3C~33.1 &30 &  0.2000 & 4 (2) & 0.2 (0.5) & 1.19             & 0.862 (316) \\
3C~61.1 & 1 & 0.0787 & $<0.04$  & 1.0 (0.15) & \nodata    & 0.989 (187) \\
3C~105 &  15 & 0.1150 & 31 (20)  & 0.20  & 1.32             & 1.091(180)  \\
3C~133  & 30 &0.2540 & 1.0 (0.4)  & 1.15 (0.3) &1.99     & 1.017(236) \\
3C~171 & 1  & 0.0565 & 6 (0.3)  & 0.20 (0.7) & \nodata      & 0.853(144) \\
3C~184.1 & 30 &0.0351 & 4.0 (1.3)  & -0.13 (0.5)  & 1.09  & 0.857 (295) \\
3C~197.1& 30 & 0.0420 & $<0.1$  & 0.4 (0.2)  & 1.30       & 0.911 (281) \\
3C~223.1 & 1 & 0.0131 & 25 (15)  & 0.98  &  \nodata           & 1.306 (131) \\
3C~287.1 & 30 & 0.0163 & $<0.03$  & 0.04 (0.1)  & 1.08  & 0.815 (328) \\
3C~293 & 1 &  0.0127 &  \nodata & \nodata  &   \nodata      & \nodata \\
3C~300  &1 &0.0249  & $<0.2$ & 0.6 (0.4)  & \nodata          & 0.816 (106) \\
3C~305  & 1  &0.0131 & \nodata & \nodata  & \nodata          & \nodata  \\
3C~323.1 & 30 &0.0379  & $<0.2$   &  -0.6 (0.2)  & 1.18   & 0.874 (282)\\
3C~332 & 30 &   0.0179 & \nodata & \nodata  & \nodata      & \nodata \\
3C~349 &1 & 0.0188  & 0.9 (0.4)  & 0.2 (0.4)  &  \nodata      & \nodata \\
3C~381 & 15 & 0.0615  & 27 (3)   & 1.0  & 0.802                   & 1.079 (178) \\
\hline
\end{tabular}\\
\end{flushleft}
\tablecomments{Summary  of  results of our spectral analysis 
for the brighter nuclei in our sample (those sources with greater than 50 counts 
in a 2\arcsec nuclear aperture). 
(1) 3C source name;
(2) bin size, in counts, of the point source spectrum extracted from
the 2\arcsec aperture centered about the nucleus.  Those sources whose
spectra have been binned to 30 (or in a few cases 15) counts were
analyzed in XSPEC using $\chi^2$ statistics.  Unbinned spectra (listed
as binned to 1 count in column 2) were analyzed using Cash statistics
(Cash 1979).  Note that all binned (30 count) spectra were also
analyzed with Cash to check for consistency in parameter fits and
model flux values. The result of this analysis showed reasonable
consistency between the two statistics models, though a
goodness-of-fit indicator was lacking for Cash statistics.
(3) Fixed galactic absorbing neutral hydrogen column density (Kalberla et al. 2005); 
(4) Best-fit redshifted neutral hydrogen absorption component. Those
sources listed with no entries have an unconstrained
N$_H\left(z\right)$, owing to (a) low intrinsic absorption and/or (b)
degeneracy in the model fit.
(5) Best-fit spectral index;
(6) reduced Chi$^2$ of the given fit, listed only for binned spectra. A goodness-of-fit indicator is 
not available for the unbinned spectra as Cash statitistics were used in their model fits.
(7) Cash statistic divided by degrees of freedom (d.o.f.).  The d.o.f. are
given in parentheses.}
\end{table*}

\subsection{Source details}\label{sec:sources}
\noindent
\underline{{\bf 3C 17}} is a broad lined radio galaxy (BLRG)
(Buttiglione et al. 2009) at a redshift of z=0.219 (Fig.~\ref{fig:17}).
The kpc scale radio morphology is dominated on the southern side by a
strongly curved jet, as described by Morganti et al. (1993, 1999).  The
northern side has lower brightness and does not have a well defined
jet.  There is also much larger, low brightness radio emission.  3C17
has been described as a ``transition object'' between FRI and FRII by
Venturi et al. (2000).  The jet has a bright knot at
3.7$^{\prime\prime}$ from the nucleus while the curved part lies at
about 11$^{\prime\prime}$.  We found X-ray and optical emission from
both these knots.  The results of the X-ray analysis and the relevance
of using the curvature in the jet as a diagnostic tool in
understanding the nature of X-ray emission in jets, has been presented
in Massaro et al. (2009b).
 
\noindent
\underline{{\bf 3C 18}} is a bright FR II BLRG (Tadhunter et al. 2002)
at redshift z= 0.188.  As
shown in Table~\ref{tab:flux}, the number of counts increased by
13$\%$ going from 2'' to 10'' radii; an indication of diffuse emission
around its nucleus.

\noindent
\underline{{\bf 3C 20}} is an FR II HEG radio galaxy with evidence for significant intrinsic absorption (see Table 2 and/or Table 4).

\noindent
\underline{{\bf 3C 33.1}} is a broad line object yet both of our
estimates for absorption show an intrinsic column density of 4 to
5~$\times10^{22}$cm$^{-2}$.  Generally we would not expect heavy
absorption for BLO's; perhaps this source is an example of absorption
with an anomolously low gas to dust ratio (see Grandi et al. 2007).

\noindent
\underline{{\bf 3C 52}} is an FR II radio galaxy and the original HST
image (de Koff et al. 2000) shows a convincing case of an obscuring
dust disk.  The radio map illustrates
that the radio axis is approximately perpendicular to the dust
disk. Deeper radio maps show that the extended radio emission of this
source is X-shaped.  In Table~\ref{tab:hotspot} we report the marginal
X-ray detection of the northern hotspot.

\noindent
\underline{{\bf 3C 105}} is a classical FRII radio galaxy.  The host
galaxy is a NLRG with only nuclear emission lines (Baum et al.  1988;
Smith \& Heckman 1989; Tadhunter et al. 1993).

We detected X-ray emission from the core and southern hotspot, both from the
location where the radio jet appears to enter the hotspot region, and
from the brightest radio emission at the terminal hotspot itself.  It
would seem that the jet, which has stretched over some 400 kpc on the
plane of the sky (undetected in both radio and X-rays), has gently
bent as it approaches the hotspot and is presumably finally much
closer to the l.o.s since the ratio of X-ray to radio intensity is
about ten times larger at this 'jet entrance' than for the hotspot
itself.  An image of this hotspot is shown in Fig.~\ref{fig:105} and a
more detailed study of 3C 105 is in preparation (Orienti et al. 2010).

Both the photometric and the spectral analsyis of the nuclear emission in 3C 105
are in agreement showing evidences of high intrinsic absorption.
For this source, there are \swf~ and \xmm~public observations,
we reduced and analyzed these data to compare our methods
with a more accurate spectral analysis.

In the \xmm~observation, we found the best fit value of the intrinsic absorption is 49.1$\pm$4.7 $\times$ 10$^{22}$ cm$^{-2}$
and also a marginal detection of an iron K$\alpha$ line with EW $\sim$ 90 eV in a
complex X-ray spectrum similar to that of 3C 33 (Torresi et al. 2009) and 3C 234 (Piconcelli et al. 2008).
The value of the intrinsic absorption found in the \swf~ data is consistent with that of \xmm, 
with N$_H$ equal to 30.4$\pm$ 5.5 $\times$ 10$^{22}$ cm$^{-2}$.
All the details on this spectral analyses are reported in the Appendix.

\noindent
\underline{{\bf 3C 132}} is an FR II LEG radio galaxy with a evidence for significant intrinsic absorption (see Table 2 and/or Table 4).

\noindent
\underline{{\bf 3C 165}}: 3C 165 is an FR II radio galaxy. We
detected only the nucleus at a position $\approx$11$^{\prime\prime}$
from the NED location.  G. Taylor (private communication) provided a
radio position for a point-like source ($\approx$~9mJy) from a recent
8.4 GHz VLA observation (J2000 6 43 6.660 + 23 19 0.55).  Our X-ray
position agrees with this so we believe it is the nucleus of the radio
galaxy.

\noindent
\underline{{\bf 3C 171}} is a small FR II radio galaxy.  It has a
slightly disturbed elliptical morphology and particularly strong
extended emission line regions.  These emission line regions are
aligned with the radio emission (Heckman et al.1984, Blundell 1996).
We detect extended X-ray emission around the nucleus consistent with a
hot gas component because there is no
one-to-one association with the radio emission, and the nucleus is
also strongly depolarized (Hardcastle et al. 2003).  We also found
X-rays associated with the western hotspot.  An analysis based on a
Chandra followup observation of 3C 171 uses radio polarization data to
interpret the extended X-ray emission (Hardcastle et al. 2009b).
We find evidence for significant intrinsic absorption (see Table 2 and/or Table 4).

\noindent
\underline{{\bf 3C 184.1}} is a broad line object yet both of our estimates
for absorption show an intrnsic column density of 4 to 5 $\times10^{22}cm^{-2}$.  
In this respect, it is similar to 3C~33.1.

\noindent
\underline{{\bf 3C 223.1}} is an FR II HEG radio galaxy with a X shape radio morphology.
In this source, we find an evidence for significant intrinsic absorption (see Table 2 and/or Table 4).

\noindent
\underline{{\bf 3C 287.1}} is a strongly core-dominated FR II
BLRG with a very similar radio structure to 3C 171. No hotspot has
been detected in the X-rays but only the bright nucleus. As in the
case of 3C18 it may be considered an example of diffuse emission
around the core.  The number of counts increases by about 11 $\%$
going from a circular region of r=2'' to 10''. 
This could be due to hot gas around the nuclear region similar to the case of 3C 18.

\begin{figure}
\includegraphics[scale=0.30,origin=c,angle=0]{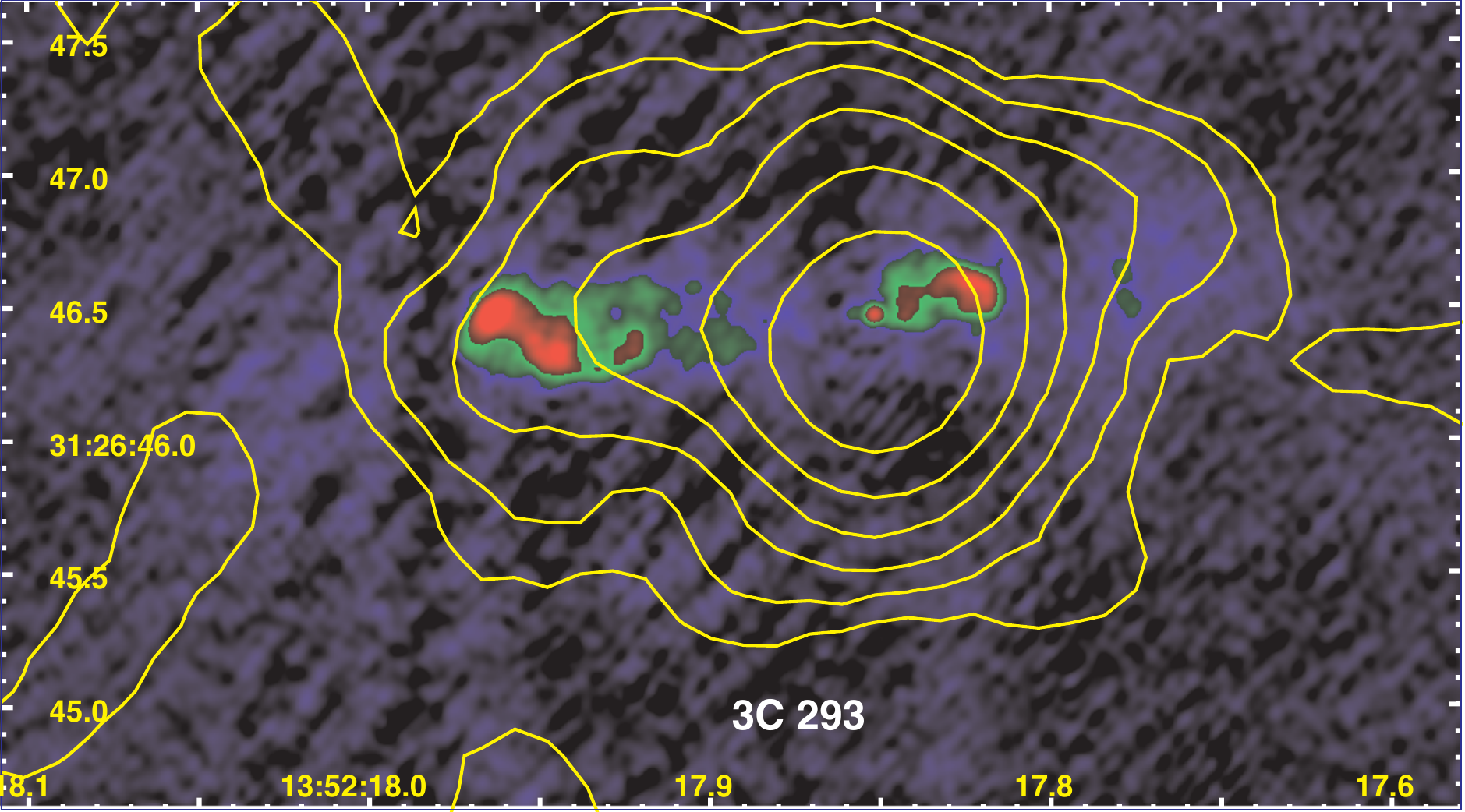}
\caption{3C~293.  A 4.86 GHz radio image from a MERLIN observation which shows the brighter, inner radio structure.
The X-ray contours come from a rebinned image as in
Fig.~\ref{fig:305}.  The smoothing function is a Gaussian of
FWHM=0.6\asec.  Contours begin at 0.05 cnts/pixel and increase by
factors of two. (Radio map available on the Merlin archive).}
\label{fig:293}
\end{figure}

\noindent
\underline{{\bf 3C 293}} has a poorly defined X-ray nucleus precluding
precise registration, due to an excess of extended emission.  Therefore, we are uncertain if the small X-ray
extension to the East of the nucleus is emission from the radio jet or
if it lies along the northern edge of the jet.  An image is shown in
Fig.~\ref{fig:293}.

VLBI observations (Giovannini et al. 2005, 2008) show a two-sided
structure with symmetric jets. The jet emission is detected on both
sides up to $\sim 20$ mas from the core, with the eastern jet being
slightly brighter. We refer to it as the main jet(J), and to the
western one as the counterjet (CJ).  This is also in agreement with
the description of Beswick et al. 2004, although their suggested jet
orientation and velocity are in contrast with the high symmetry of the
VLBA jet.  From a comparison between different images (Giovannini et
al. 2008) we find that the sub-arcsecond structure in the E-W direction
is clearly related to a restarted activity of the central AGN, however
the large change in PA (position angle) with respect to the extended lobes is not due
to a PA change in the restarted nuclear activity but it looks constant
in time and it is likely produced by the jet interaction with a
rotating disk as discussed in van Breugel et al. 1984. In this
scenario we expect that the jet at $\sim$ 2.5 kpc from the core is no
longer relativistic.
We find evidence for significant intrinsic absorption (see Table 2 and/or Table 4).

\noindent
\underline{{\bf 3C 303.1}}  O'Dea et al. (2006)
analyzed XMM data and suggested that the spectrum was consistent with a 
contribution from hot gas which had
been shocked by the radio source. The Chandra data show X-ray emission 
elongated along the radio source axis
and with a size comparable to that of the radio source consistent with 
this interpretation.
The source is very small and it is hard to separate the nuclear component from the extended emission.

\noindent
\underline{{\bf 3C 305}} is a relatively low power radio galaxy
located at redshift $z$ $\sim$ 0.0416 (see Fig.~\ref{fig:305}). It has
been classified as a peculiar FR I on the basis of the radio power and
morphology (Heckman et al. 1982) and it presents a prominent extended
emission line region of roughly the same dimension as the radio
structure.  It is one of the more interesting sources in our sample,
because it is an FR~I - CSS source in which the X-ray emission is not
coincident with the radio, but there is clear evidence for an
association between the X-ray emission and [O III]5007.
A detailed comparison of the radio, optical and X-ray images (here reported) 
of 3C 305 is used to illucidate the nature of
the emitting gas in Massaro et al. (2009a).

\noindent
\underline{{\bf 3C 323.1}} is a quasar with a bright nucleus in the
X-rays. The radio morphology is that of a classical FR II but no
hotspots have been detected in the X-ray band.  Similar to the cases
of 3C 18 and 3C 287.1 this source is a good example of diffuse
emission around the nucleus.  The number of counts increases by 25
$\%$ going from a circular region of r=2'' to r=10''.

\noindent
\underline{{\bf 3C 332}} is a powerful FR II radio galaxy with a
prominent quasar-like infrared nucleus marking the center of an
elliptical host galaxy. This source has a clear detection of diffuse
emission around the nucleus.
The number of counts increases by 42$\%$ going from a circular region of r=2'' to r=10''
both centered on the radio position of the core.
We find evidence for significant intrinsic absorption (see Table 2 and/or Table 4).

\noindent
\underline{{\bf 3C 349}} is a FR II radio galaxy that lies at redshift
z=0.205.  We detect the southern hotspot and there appears to be two
components, so it might be similar to the S hotspot of 3C~105.

\noindent
\underline{{\bf 3C 381}} is an FR II HEG radio galaxy 
with an evidence for significant intrinsic absorption (see Table 2 and/or Table 4).

\noindent
\underline{{\bf 3C 436}} is an FR II HEG radio galaxy 
with an evidence for significant intrinsic absorption (see Table 2 and/or Table 4).

\noindent
\underline{{\bf 3C 460}} is not well registered since the only available (to us) radio map 
does not detect the nuclear emission, however the core is clearly detected in the X-rays, as shown in Fig. 35.  
There is a single event coincident with the N hotspot, and 2 counts near the S (brighter in radio) hotspot.

\section{Summary}\label{sec:summary}
We have presented our analyses of the Chandra 3C Snapshot Survey.  
Our goal is to obtain X-ray data for all extragalactic 3C sources with
z$<$0.3 so as to have a complete, unbiased sample.  Our AO9 proposal
resulted in the current new data for 30 of the (then) unobserved 60
sources not already in the archive with exposures of 8ks or greater.
We will submit a Chandra proposal for AO12 in order to complete the
sample.
When the remaining 27 are observed, it will then be possible 
to have a complete sample by
recovering many more 3C sources from the archives.

We have constructed flux maps for all the observations and given
photometric results for the nuclei and radio hotspots.  For the
stronger nuclei, we have employed the usual spectral analysis, and
compared the column densities of intrinsic absorption to those
obtained from the ratio of hard to medium fluxes.  As expected, we
find a sizable fraction (1/3) of our sources showing evidence for
significant absorbtion (N$_H~>~5\times~10^{22}$~cm$^{-2}$).
In particular, for 3C~105, we analyzed the archival observations performed by \xmm~and \swf~
and we found that its core is absorbed and these spectra are in agreement with our
Chandra {\bf spectral} analysis. We also found a marginal detection of the Fe K$\alpha$ emission line
and the evidence of a soft X-ray excess.

We have found several sources worthy of more detailed study, and with
collaborators have published papers on 3C~17, 3C~171, and 3C~305.  Our
decision to waive proprietary rights has permitted others (e.g. Hardcastle et al. 2009) to use
relevant data in a timely manner, and we intend to continue this
policy in our Chandra AO12 proposal for the remaining unobserved
sources.

\acknowledgments 

We thank the anonymous referee for useful comments that led to improvements in the paper.
We are extremely grateful to S. Bianchi for his help in the \xmm~data reduction procedure
and analysis of the X-ray spectrum of 3C 105 and to M. L. Conciatore for 
her suggestions in the use of \sx~data for the same source.
We thank R. Morganti for giving us her 5 GHz VLA map of 3C~17,
R. Beswick for that of 3C~293 and A. Tilak for the HST images of 3C 171.  
We are grateful to M. Hardcastle for providing several radio
maps of the 3C sources.  This research has made use of NASA's
Astrophysics Data System; SAOImage DS9, developed by the Smithsonian
Astrophysical Observatory; and the NASA/IPAC Extragalactic Database
(NED) which is operated by the Jet Propulsion Laboratory, California
Institute of Technology, under contract with the National Aeronautics
and Space Administration.  Several radio maps were downloaded from the
NVAS (NRAO VLA Archive Survey).  The National Radio Astronomy
Observatory is operated by Associated Universities, Inc., under
contract with the National Science Foundation.  
Several radio maps have been provided by the Merlin archive.
The work at SAO is supported by NASA-GRANT GO8-9114A and the work at RIT was supported by
Chandra grant GO8-9114C.
F. Massaro acknowledges the Foundation BLANCEFLOR Boncompagni-Ludovisi, n'ee Bildt 
for the grant awarded him in 2009.

{\bf Facilities:} \facility{VLA}, \facility{Merlin}, \facility{HST}, \facility{CXO (ACIS)},\\ \facility{XMM (EPIC PN)}, \facility{SWIFT (XRT)}

\appendix

\section{Images of the sources}

Although for many of our sources the X-ray data are comprised of
rather few counts, we show here the radio mophology via contour
diagrams which are superposed on X-ray fluxmaps that have been
smoothed with a Gaussian.  The full width half maximum (FWHM) of the
Gaussian smoothing function is given in the figure captions and varies
between 0.25$^{\prime\prime}$ and 2$^{\prime\prime}$.  When there is
sufficient S/N (signal to noise ratio) of the X-ray image to provide spatial information, we
have added contours (cyan or white)
which are normally separated by factors of two.  Most of the overlayed
radio contours increase by factors of four.  Our standard X-ray
fluxmaps have been rebinned by a factor of 4 (``f4'') to give a pixel
size of 0.123$^{\prime\prime}$.  For radio sources of large angular
extent, we occasionally use f2 instead of f4 in order that our
standard array size of 1024 covers the entire source.  Brightness
units of the flux maps are ergs~cm$^{-2}$~s$^{-1}$~pixel$^{-1}$.
The primary reason figures appear so different from each other is the
wide range in angular size of the radio sources.  

\begin{figure}
\includegraphics[keepaspectratio=true,scale=0.90]{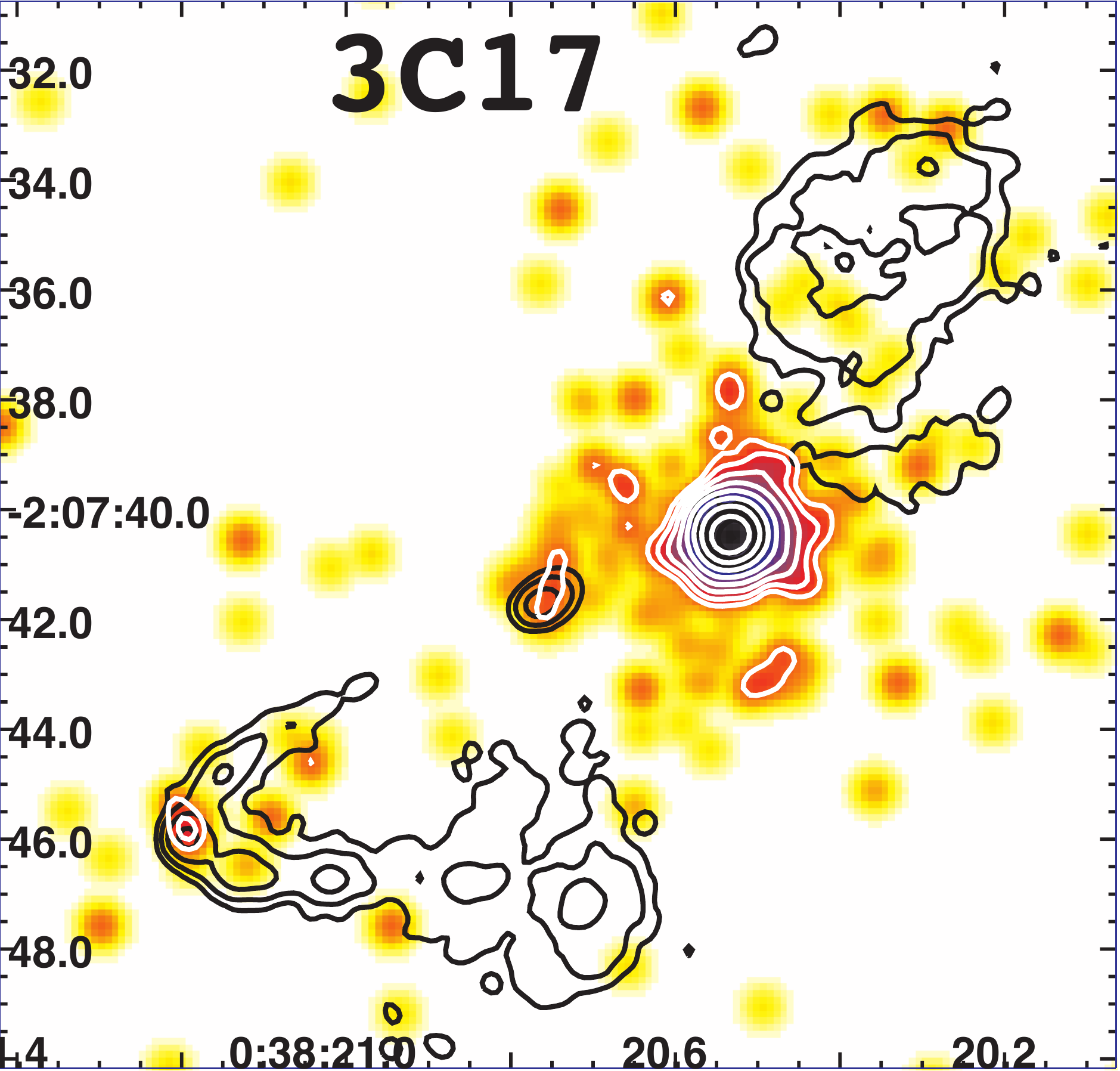}
\caption{3C17.  The Chandra flux map (0.5-7keV) has been rebinned by a
  factor of 4 (``f4'') and smoothed with a Gaussian of
  FWHM=0.5$^{\prime\prime}$.  X-ray contours start at
  1$\times10^{-16}$ and increase by factors of 2.  The 5 GHz radio map
  is from the VLA and has a restoring beam with FWHM=0.4$^{\prime\prime}$.
  Radio contours start at 0.5mJy/beam and increase by factors of 4.
  (Radio map kindly supplied by R. Morganti).}
\end{figure}

\clearpage

\section{Spectral analysis of the nuclear emission in 3C 105}
The data reduction has been performed following the procedures described 
in the Massaro et al. (2008a,b) and Bianchi et al. (2009)
both for the SWIFT and XMM-Newton data, here we report 
only the basic parameters used on the data reduction.

\subsection{\swf~observations}
3C~105 has been observed by \swf~ 4 times between July 11 and July 16, 2006 (Obs ID 00035625001-2-3-4) ,
operating with all the instruments in data taking mode 
and with an exposure of about 7 ks, 3 ks, 4 ks, 5ks, respectively.
We added all these observation to increase the S/N ratio.

For our analysis we considered only \sx~ (Burrows et al. 2005) data.
The \sx~data analysis has been performed with the XRTDAS 
software distributed within the HEAsoft package (v.~6.8.0). 
Event files were calibrated and cleaned 
with standard filtering criteria using the \textsc{xrtpipeline} task, 
combined with the latest calibration 
files available in the \swf~ CALDB distributed by HEASARC. 
Events in the energy range 0.3--10 keV with grades 0--12 
(photon counting (PC) mode) were used in the analyses 
(see Hill et al.  2004 for a description of readout modes).
No signatures of pile-up were found in our SWIFT observations.
Events are extracted using a 20 pixel radius circle. 
The background for PC mode is estimated from a nearby 
source-free circular region of 20 pixel radius.

\subsection{\xmm~observations}
3C 105 was observed with \xmm~(ObS ID 0500850401) on Aug 25, 2008 with
all EPIC CCD cameras: the EPIC-PN (Struder et al. 2001), 
and EPIC-MOS  (Turner et al. 2001), operating in Full frame mode and
with Thin filter.
Only EPIC-PN data are reported in this work, the EPIC-MOS data 
have been reduced and analyzed and we found them in agreement
with our results.

Extractions of all light curves, source and background spectra  are done
using the \xmm~ Science  Analysis System (SAS) v9.0.0.  The calibration
index file  (CIF) and  the summary file  of the observation data file
(ODF) were  generated using updated calibration files (CCF) following the ``User's Guide  to
the \xmm~ Science Analysis System"  (issue 3.1, Loiseau et  al. 2004) and ``The \xmm~
ABC Guide"  (vers. 2.01, Snowden et al. 2004). 
Event files were produced by the \xmm~ EPCHAIN pipeline.

Photons are extracted from a circular region. The
typical value of the external radius for the circular  region 
is $14$~$^{\prime\prime}$.
A restricted energy range (0.5--10 keV) is used to avoid
possible residual calibration uncertainties.   

To ensure the validity of  $\chi^2$  statistics, both spectra, 
\swf~ and \xmm, are  grouped  by  combining
instrumental channels so that each new bin comprises  30 counts or more, 
well above the limit for the $\chi^2$ test applicabilty (Kendall \& Stuart, 1979).

\subsection{Spectral analysis}
We initially fitted the \xmm~spectrum in the energy range 2-10 keV with a model composed by: the galactic absorption (see Tab.1), 
a power-law (spectral index $\alpha$ = 0.83 $\pm$ 0.25), an intrinsic absorption ($N_H^{int}$= 49.1 $\pm$ 4.7 10$^{22}$ cm$^{-2}$)
plus an intrinsic narrow iron K$\alpha$ emission line (gaussian profile, 
at energy $E_{K\alpha}$ = 6.36 $\pm$ 0.05 keV). 
In Fig. 36 the \xmm~spectrum in the 2-10 keV energy range is shown, the $\chi^2$ for the model proposed is 36.7 over 38 d.o.f.
We also found the evidence of a soft excess at energies below 2 keV, that could be due to a blend 
of emission lines as the case of other radio galaxies (e.g. 3C 33, Torresi et al. 2009 and 3C 234, Piconcelli et al. 2008) or it could be interpreted as 
soft X-ray emission from the inner jet of the host radio galaxy.
Finally, both \sx~and \xmm~spectral analysis are in agreement indicating evidences of intrinsic absorption,
however the low S/N ratio of the \sx~observation does not allow a good estimate of its value.
The photometric and the spectral method in this case are not in agreement because the spectral analysis revealed a more detailed spectrum than a simple featureless power-law continuum.

\begin{figure}
\begin{center}
\includegraphics[keepaspectratio=true,scale=0.35, angle=-90]{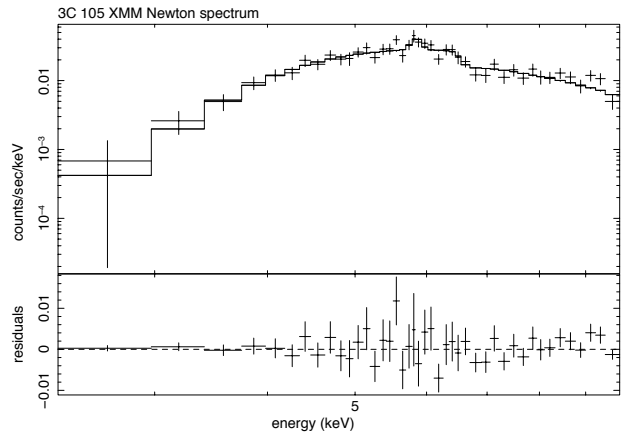}
\caption{The \xmm~ EPIC-PN spectrum of the nuclear emission in 3C 105 with its residuals, fitted with the bestfit model.}
\end{center}
\end{figure}

\vspace{4cm}
\begin{center}
{\bf THE COMPLETE VERSION OF THE PAPER, INCLUDING ALL 30 FIGURES IN THE APPENDIX,  IS AVAILABLE AT \\
\vspace{0.4cm}
\underline{http://hea-www.harvard.edu/~harris/3c/}\\
\vspace{0.4cm}
BOTH IN PDF AND PS VERSION.}
\end{center}
\vspace{4cm}

\end{document}